\documentclass[12pt]{article}
\usepackage{amssymb}
\usepackage{amsfonts}
\usepackage{amsmath}
\usepackage{graphicx}
\usepackage{color}
\usepackage{booktabs}
\usepackage{multirow}
\usepackage{enumerate}
\usepackage{hyperref}

\hypersetup{colorlinks=true, linkcolor=blue}

\newtheorem{theorem}{{\bf \sc Theorem}}
\newtheorem{lemma}{{\bf \sc Lemma}}

\newtheorem{corollary}{{\bf \sc Corollary}}

\input{tcilatex}
\begin{document}

\title{Diffusion and Contagion in Networks with Heterogeneous Agents and
Homophily\thanks{%
The first author acknowledges support from the NSF under grant SES--0961481.
The second author acknowledges support from the Spanish Ministry of Science
and Innovation (ECO2008-03883, ECO2011-22919) as well as from the Andalusian
Department of Economy, Innovation and Science (SEJ-4154, SEJ-5980) via the
\textquotedblleft FEDER operational program for Andalusia,
2007-2013\textquotedblright .}}
\author{Matthew O. Jackson\thanks{%
Stanford University, Santa Fe Institute, and CIFAR} \ and Dunia L\'{o}%
pez-Pintado\thanks{%
Universidad Pablo de Olavide and CORE, Universit\'{e} catholique de Louvain.}%
}
\date{Draft: October, 2011}
\maketitle

\begin{abstract}
We study how a behavior (an idea, buying a product, having a disease,
adopting a cultural fad or a technology) spreads among agents in an a social
network that exhibits segregation or homophily (the tendency of agents to
associate with others similar to themselves). Individuals are distinguished
by their types (e.g., race, gender, age, wealth, religion, profession, etc.)
which, together with biased interaction patterns, induce heterogeneous rates
of adoption. We identify the conditions under which a behavior diffuses and
becomes persistent in the population. These conditions relate to the level
of homophily in a society, the underlying proclivities of various types for
adoption or infection, as well as how each type interacts with its own type.
In particular, we show that homophily can facilitate diffusion from a small
initial seed of adopters.


\end{abstract}

\section{Introduction}

Societies exhibit significant homophily and segregation patterns.\footnote{%
For background on homophily and some of its consequences, see McPherson et
al. (2001) and Jackson (2008).} How do such biases in interactions affect
the adoption of products, contagion of diseases, spread of ideas, and other
diffusion processes? For example, how does the diffusion of a new product
that is more attractive to one age group depend on the interaction patterns
across age groups? How does the answer depend on the differences in
preferences of such groups, their relative sociabilities, and biases in the
interactions?

We answer these questions by analyzing a general model of diffusion that
incorporates a variety of previous models as special cases, including
contagion processes studied in the epidemiology literature such as the
so-called SIS model (e.g., Bailey 1975, Pastor-Satorr\'{a}s and Vespignani,
2001), as well as interactions with strategic complementarities, such as in
the game theoretic literature and network games (e.g., Galeotti et al.,
2010).\footnote{%
For background on diffusion in networks see Newman (2002), Jackson and Yariv
(2005, 2007, 2010), L\'{o}pez-Pintado (2006, 2008, 2010), Jackson and Rogers
(2007) among others.} Our model incorporates types of individuals who have
different preferences or proclivities for adoption, as well as biases in
interactions across types.

In particular, we examine whether or not diffusion occurs from a very small
introduction of an activity in a heterogeneous and homophilous society. We
first concentrate on the focal situation with only two types of agents.
Within this case, the most interesting scenario turns out to be one where
one type would foster diffusion and the other would not if the types were
completely segregated. In that scenario, we show that homophily actually
facilitates diffusion, so that having types biased in interactions towards
their own types can enhance diffusion to a significant fraction of both
types. Having a higher rate of homophily, so that a group is more
introspective, allows the diffusion to get started within the group that
would foster diffusion on its own. This can then generate the critical mass
necessary to diffuse the behavior to the wider society. In contrast,
societies exhibiting less homophily can fail to foster diffusion from small
initial seeds.

We then move to the general case of many types. Our main characterization
theorem generalizes the features from the two-agent case, showing that
diffusion relates to a condition on the largest eigenvalue of an interaction
matrix which tracks the initial adoption rates of various types of
individuals, that is, their adoption rates from small initial seeds. Again,
we show that homophily can facilitate diffusion, showing that a sufficient
condition is that some type (or group of types) that would adopt on its own
is sufficiently homophilous to give the diffusion a toehold. We discuss how
this extends the intuitions from the case of two types.


\section{An Illustrative Example: The Heterogeneous SIS model with Two Types 
\label{sectSIS}}

To fix ideas and preview some of the insights from the general model, we
begin with a case where there are just two types of agents and the contagion
follows a simple and well-studied process.

In particular, consider an infectious disease spreading in a population with
two groups: the young and the old. Our aim is to analyze whether or not {%
diffusion} of the disease occurs. That is, if we start with a small seed of
infected agents, will the infection spread to a significant fraction of both
populations and become endemic? In order to answer this question consider
the following {heterogeneous} version of the canonical SIS model.\footnote{%
The so-called SIS (Susceptible-Infected-Susceptible) model is a basic one
used by the epidemiology literature to describe such situations (e.g.,
Bailey 1975, Past\'{o}r-Satorr\'{a}s and Vespignani, 2000, 2001).}

Agents can be in one of two \textquotedblleft states\textquotedblright :
infected or susceptible. A susceptible agent becomes infected at an
independent probability $\nu >0$ from each interaction with an infected
agent. Conversely, with a probability $\delta >0$ per unit of time an
infected individual recovers and becomes susceptible again.\footnote{%
The SIS model allows a recovered person to catch the disease again. An
obvious instance is the standard flu.} The crucial parameter of the model is
the relative spreading rate, $\lambda =\frac{\nu }{\delta }$, which measures
how infectious the disease is in terms of how easy it is to contract
compared to the rate at which one recovers.

An interesting case for our analysis is one where the population is
heterogeneous in terms of the proclivities for getting infected. In
particular, imagine that the older are more (or less) vulnerable to the
disease than the young. More precisely, if $\lambda _{1}$ is the spreading
rate of the young and $\lambda _{2}$ of the old, then we allow $\lambda
_{1}\neq \lambda _{2}$.

In addition to their age, individuals are also potentially differentiated by
the rates at which they interact with other individuals, where ``interact''
is taken to mean that they have a meeting with an individual which could
transmit the infection if one of them is infected and the other is
susceptible. In particular, apart from his or her type, each individual is
characterized by a degree $d$; the number of agents the individual meets
(and is potentially infected by) every period. Let $P_{i}(d)$ be the degree
distribution of individuals of type $i$; that is, the fraction of agents of
type $i$ that have $d$ meetings per unit of time.

Also, for the purposes of this example, we stick with what is standard in
the random network literature, and take the meeting process to be
proportionally biased by degree. Thus, conditional on meeting an agent of
type $i$, the probability that he or she is of degree $d$ is $\frac{d}{%
\langle d\rangle _{i}}$, where $\langle d\rangle _{i}$ is the average degree
among type $i$ agents ($\langle d\rangle _{i}=\sum_{d}P_{i}(d)d$).

To capture homophily, let $0<\pi<1$ be the probability that a given type $i$
agent (old or young) meets his or her own type, and $1-\pi$ be the
probability of meeting an agent of the other type. For example if the
populations are of even size, then having $\pi>1/2$ means that agents are
mixing with their own type disproportionately.

We say that diffusion occurs from a small seed (with a formal definition
below) if starting from an arbitrarily small amount of infected individuals
(of either type), we end up with a nontrivial steady-state infection rate
among the population.

Let $\pi _{0}=\frac{1-\widetilde{d}_{1}\lambda _{1}\widetilde{d}_{2}\lambda
_{2}}{\widetilde{d}_{1}\lambda _{1}+\widetilde{d}_{2}\lambda _{2}-2%
\widetilde{d}_{1}\lambda _{1}\widetilde{d}_{2}\lambda _{2}}$, where $%
\widetilde{d}_{i}=\frac{\langle d\rangle _{i}^{2}}{\langle d\rangle _{i}}$.

\begin{theorem}
\label{the-one} Diffusion occurs from a small seed in the two type SIS model
if and only if one of the following holds:

1) $\lambda _{1}\lambda _{2}>\frac{1}{\widetilde{d}_{1}\widetilde{d}_{2}}$ or

2) $\lambda _{1}\lambda _{2}<\frac{1}{\widetilde{d}_{1}\widetilde{d}_{2}}$
and $\pi >\pi _{0}$
\end{theorem}

The proof of the theorem appears in the appendix, and is a special case of
our more general results below.

The condition for diffusion in the standard (homogeneous) SIS model is $%
\lambda >\frac{1}{\widetilde{d}}$ (e.g., Pastor-Satorr\'{a}s and Vespignani,
2001). Thus, we see how this generalizes in the above theorem.

Theorem \ref{the-one} yields the following straightforward consequences.

\begin{corollary}
\label{cor-one} The following statements hold for the two-type SIS model:

1) If diffusion occurs within each type when isolated (when $\pi=1$), then
it would also occur when there is interaction among the two (when $\pi<1$).

2) If diffusion does not occur in either of the types when isolated, then it
would not occur when there is interaction among the two.

3) If diffusion occurs among one type but not the other when isolated, then
it will occur among the whole population if the homophily is high enough.
\end{corollary}

The most interesting scenario turns out is the last one, such that one of
the types would foster diffusion if isolated, whereas the other would not
(i.e., $\lambda _{1}>\frac{1}{\widetilde{d}_{1}}$ and $\lambda _{2}<\frac{1}{%
\widetilde{d}_{2}}$). In that scenario, homophily either plays no role (that
is, when $\lambda _{1}\lambda _{2}>\frac{1}{\widetilde{d}_{1}\widetilde{d}%
_{2}}$) so that any homophily level will allow diffusion, or else it
actually facilitates diffusion (that is, when $\lambda _{1}\lambda _{2}<%
\frac{1}{\widetilde{d}_{1}\widetilde{d}_{2}}$ in which case $\pi $ must
exceed $\pi _{0}$).

In the latter case diffusion occurs only if the two types are sufficiently
biased in interactions towards their own types (i.e., $\pi $ is sufficiently
large). The intuition for such a result is the following. Having a higher
rate of homophily, so that a group is more introspective, allows the
diffusion to get started within the group that would foster diffusion on its
own. In turn, it can then spread to the wider society.

\section{The General Model}

With this introduction behind us, we now describe the general model.

\subsection{Types and Degrees}

Each agent is characterized by his or her degree $d\geq 0$ and type $i\in
T=\{1,...,m\}$.

Since the number of individuals of each type can differ, let $n(i)$ be the
fraction of individuals of type $i$.

An agent's degree $d$ indicates the number of other agents that the agent
meets (and is potentially influenced by) before making a decision in a given
period. The meeting process is allowed to be directional; i.e., agent $h$
meeting (paying attention to) agent $k $ does not necessarily imply that $k$
pays attention to $h$. So, although we use the term ``meeting,'' the
interaction need not be reciprocal. Of course, a special case is one where
the interaction is mutual.

Different types may have different distributions in terms of how frequently
they meet other agents. In particular, let $P_{i}(d)$ be the degree
distribution of individuals of type $i$. That is, $P_{i}(d)$ is the fraction
of type $i$ individuals who have $d$ meetings per period. Thus, there can be
heterogeneity among agents of a given type, in terms of how social they are.

An agent's type $i$ shapes both the agent's relative interaction rates with
other types of agents and the agent's preferences or proclivity for
infection. In particular, $\pi _{ij}$ is the probability that an agent of
type $i$ meets an agent of type $j$ in any given meeting. Clearly, $%
\dsum\limits_{j=1}^{m}\pi _{ij}=1$. The bias in meetings across types is
then summarized by the matrix 
\begin{equation*}
\Pi =\left( 
\begin{array}{ccc}
\pi _{11} & \ldots & \pi _{1m} \\ 
\vdots & \ldots & \vdots \\ 
\pi _{m1} & \ldots & \pi _{mm}%
\end{array}%
\right).
\end{equation*}

We assume that $\Pi$ is a primitive matrix (so that $\Pi^t>0$ for some $t$).
This ensures that there is at least some possibility for an infection that
starts in one group to reach any other, as otherwise there are some groups
that are completely insulated from some others.

\subsection{The Random Meeting Process}

In order to study this system analytically, we examine a continuum of
agents, $N=[0,1]$.

This continuum is partitioned into agents of different types, and then
within types, by their degrees.

There are two ways in which the meeting process can be biased: by type and
by degree.

In particular, as mentioned above, the relative proportion of a type $i$
agent's meetings with type $j$ is described by the term $\pi_{ij}$, which
captures relative biases in meetings across types. So, in a given period, an
agent of type $i$ with degree $d$ expects to meet $d\pi_{ij}$ agents of type 
$j$. Those agents are randomly selected from the agents among type $j$.

We also allow the meeting process to be biased by degree. The probability
that an agent meets an agent of degree $d$ out of those of type $j$ is given
by 
\begin{equation*}
P_{j}(d)w_{j}(d).
\end{equation*}%
If there is no weighting by degree, then an agent equally samples all agents
of type $j$ and $w_{j}(d)=1$. This would require a directed meeting process,
such that an agent observes members of a given type uniformly at random,
independently of their meeting process or sociability. If instead, meetings
are proportional to how social the agents of type $j$ are, then $%
w_{j}(d)=d/\langle d\rangle _{j}$, where $\langle d\rangle _{j}$ is the
average degree among type $j$ agents. This latter condition covers cases in
which meetings are reciprocal.\footnote{%
For some details and references for random meeting processes on a continuum,
see the appendix of Currarini, Jackson and Pin (2009).}

Our formulation also allows for other cases. For simplicity, we assume that $%
w_{j}(d)> 0$ for all $j$ and $d$ such that $P_{j}(d)> 0$.

\subsection{The Infection or Adoption Process}

In each period an agent is in one of two states $s\in \{0,1\}$. Either the
agent has adopted the behavior and are in state $s=1$ (active, adopted,
infected...), or they have not adopted the behavior and are in state $s=0$
(passive, non-adopter, susceptible...). The agents' actions are influenced
by the actions of others, but in a stochastic manner.

Agents are heterogeneous with respect to their proclivities to adopt the
behavior. A passive agent of type $i$ adopts the behavior at a rate
described by a function $f_{i}(d,a)$, where $d$ is the agent's degree
(number of meetings per unit of time) and $a$ is the number of agents whom
she meets who have adopted the behavior. (Details of the dynamics will be
given below.) The reverse process, by which an active agent of type $i$
becomes passive happens at a rate described by a function $g_{i}(d,a)$. The
functions $f_{i}(d,a)$ and $g_{i}(d,a)$ are the primitives of the diffusion
process and are assumed to satisfy some basic conditions:

\begin{description}
\item[A1] $f_{i}(d,0)=0$ for each $i$ and $d$: a passive agent cannot become
active unless she meets at least one active agent.

\item[A2] $f_{i}(d,a)$ is non-decreasing function in $a$: the adoption rate
is non-decreasing in the number of active agents met.

\item[A3] $f_{i}(d,1)>0$ for each $i$ and some $d$ such that $P_{i}(d)>0$.
This condition implies that for each type of agent there exists some degree
such that the rate of adoption for agents with such a degree is positive
when they meet at least one active agent.

\item[A4] $g_{i}(d,0)>0$ for each $i$ and $d$: it is possible to return from
active to passive when all agents met are passive.

\item[A5] $g_{i}(d,a)$ is non-increasing in $a$: the transition rate from
active to passive is non-increasing in the number of active agents met.
\end{description}

This general model of diffusion admits a number of different models,
including models based on best-response dynamics of various games (with
trembles) as well as epidemiological models. Here are a few prominent
examples of processes that are admitted:

\begin{itemize}
\item Susceptible-Infected-Susceptible (\textbf{SIS diffusion process}): $%
f_{i}(d,a)=\nu _{i}a$ and $g_{i}(d,a)=\delta _{i}$, where $\nu _{i}\geq 0$
and $\delta _{i}\geq 0$.

\item Myopic-best response dynamics by agents who care about the \textsl{%
relative} play of neighbors (\textbf{Relative Threshold diffusion process}): 
$f_{i}(d,a)=\nu _{i}$ if $\frac{a}{d}\geq q$ and $f_{i}(d,a)=0$ otherwise.
Also $g_{i}(d,a)=\delta _{i}$ if $\frac{a}{d}<q$ and $g_{i}(d,a)=0$
otherwise, where $\nu _{i}\geq 0$ and $\delta _{i}\geq 0$ and $q\in \lbrack
0,1]$.

\item Myopic-best response dynamics by agents who care about the \textsl{%
aggregate} play of neighbors (\textbf{Aggregate Threshold diffusion process}%
): $f_{i}(d,a)=\nu _{i}$ if $a\geq \min [q,d]$ and $f_{i}(d,a)=0$ otherwise.
Also, $g_{i}(d,a)=\delta _{i}$ if $a<q$ and $g_{i}(d,a)=0$ otherwise, where $%
\nu _{i}\geq 0$ and $\delta _{i}\geq 0$ and $q\geq 0$.\footnote{%
In order to satisfy [A3] in this case, it is necessary to have some
probability of degree 1 agents for each type, or else to have $q=1$.}

\item Imitation dynamics when a neighbor is chosen uniformly at random (%
\textbf{Imitation diffusion process}): $f_{i}(d,a)=\nu _{i}\frac{a}{d}$ and $%
g_{i}(d,a)=\delta _{i}(1-\frac{a}{d})$ , where $\nu _{i}\geq 0$ and $\delta
_{i}\geq 0$.
\end{itemize}

\subsection{Steady States and Dynamics}

In order to keep track of how diffusion or infection occurs, we analyze a
continuous time dynamic, where at any given time $t\geq 0$ the state of the
system consists of a partition of the set of agents in ``active'' and
``passive.''

As is standard in the literature, we study the continuous system as an
analytically tractable alternative to the stochastic discrete system.%
\footnote{%
See Jackson (2008) for discussion of what is known about the approximation.}

Let $\rho _{i,d}(t)$ denote the frequency of active agents at time $t$ among
those of type $i$ with degree $d$. Thus, 
\begin{equation*}
\rho _{i}(t)=\sum\limits_{d}P_{i}(d)\rho _{i,d}(t)
\end{equation*}%
is the frequency of active agents at time $t$ among those of type $i$, and 
\begin{equation*}
\rho (t)=\sum\limits_{i} n(i)\rho _{i}(t)
\end{equation*}%
is the overall fraction of active agents in the population at time $t$.

The adoption dynamics are described as follows:%
\begin{equation}
\frac{d\rho _{i,d}(t)}{dt}=-\rho _{i,d}(t)rate_{i,d}^{1\rightarrow
0}(t)+(1-\rho _{i,d}(t))rate_{i,d}^{0\rightarrow 1}(t)\text{,}
\label{rho-id}
\end{equation}%
where $rate_{i,d}^{0\rightarrow 1}(t)$ is the rate at which a passive agent
of type $i$ and with degree $d$ becomes active, whereas $rate_{i,d}^{0%
\rightarrow 1}(t)$ stands for the reverse transition. In order to compute
these transition rates we must calculate first the probability that an agent
of type $i$ has of sampling an active agent. Denote this probability by $%
\widetilde{\rho }_{i}(t)$. It is straightforward to see that 
\begin{equation}  \label{two}
\widetilde{\rho }_{i}(t)=\sum\limits_{j}\pi
_{ij}\sum\limits_{d}P_{j}(d)w_{j}(d)\rho _{j,d}(t).
\end{equation}

Given $\widetilde{\rho }_{i}(t)$ then 
\begin{equation*}
rate_{i,d}^{0\rightarrow 1}(t)=\sum_{a=0}^{d}f_{i}(d,a)\left(
_{a}^{d}\right) \widetilde{\rho }_{i}(t)^{a}(1-\widetilde{\rho }%
_{i}(t))^{(d-a)}
\end{equation*}%
and%
\begin{equation}
rate_{i,d}^{1\rightarrow 0}(t)=\sum_{a=0}^{d}g_{i}(d,a)\left(
_{a}^{d}\right) \widetilde{\rho }_{i}(t)^{a}(1-\widetilde{\rho }%
_{i}(t))^{(d-a)}.  \label{rates}
\end{equation}

A \textsl{steady-state} is when $\frac{d\rho _{i,d}(t)}{dt}=0$, which
implies that we can write the steady state level $\rho _{i,d}(t)$ as being
independent of time. Solving from equation (\ref{rho-id}) leads to the
following necessary condition 
\begin{equation}
\rho _{i,d}=\frac{rate_{i,d}^{0\rightarrow 1}}{rate_{i,d}^{0\rightarrow
1}+rate_{i,d}^{1\rightarrow 0}}\text{.}  \label{rho-equilibrium}
\end{equation}

If we specify the rates $\widetilde{\rho }_{i}(t)$ for each type $i$, then
this determines the rates of transition under (\ref{rates}). This in turn,
leads to a level of $\rho _{i,d}$ for each $i,d$ under (\ref{rho-equilibrium}%
) that would have to hold in equilibrium, which in turn determines the rates
at which active agents would be met under $\widetilde{\rho }_{i}(t)$. Thus,
replacing equation (\ref{rho-equilibrium}) in equation (\ref{two}) we find
that a steady state equilibrium corresponds to a fixed point calculation as
follows: 
\begin{equation}
\widetilde{\rho }_{i}=H_{i}(\widetilde{\rho }_{1}\ldots,\widetilde{\rho }%
_{n}),  \label{equ}
\end{equation}%
where 
\begin{equation*}
H_{i}(\widetilde{\rho }_{1},\ldots,\widetilde{\rho }_{n})=\sum\limits_{j}\pi
_{ij}\sum\limits_{d}P_{j}(d)w_{j}(d)\frac{rate_{j,d}^{0\rightarrow 1}}{%
rate_{j,d}^{0\rightarrow 1}+rate_{j,d}^{1\rightarrow 0}}
\end{equation*}

The previous system of equations implicitly characterizes the steady states
of the dynamics, since by solving for $\widetilde{\rho }_{i}$ we can easily
find the fraction of adopters of each type $\rho _{i}$ and ultimately the
overall fraction of adopters $\rho $.

\subsection{Diffusion or Contagion from a Small Seed}

We now consider the following question which is the central focus of our
analysis: If we start with a small fraction of adopters, would the behavior
spread to a significant fraction of the population(s)? In other words, we
determine the conditions that lead to the diffusion of a new behavior to a
significant fraction of the population when there is a small initial
perturbation of an initial state in which nobody is infected or has adopted
the behavior; so starting from $\mathbf{(}\rho _{1},\ldots, \rho _{n}\mathbf{%
)}=(0,\ldots,0)$.\footnote{%
Notice that the question of moving away from all 1 is completely analogous,
simply swapping notation between 0 and 1 throughout the model.}

Thus, in what follows we explore the behavior of the system of (\ref{equ})
near $\widetilde{\mathbf{\rho }}=\overrightarrow{0}$; in order to see
conditions under which it is a \textsl{stable} steady-state.

The system of equations described in (\ref{equ}) can be approximated by a
linear system in the neighborhood of $\widetilde{\mathbf{\rho }}=%
\overrightarrow{0}$ as follows: 
\begin{equation*}
\widetilde{\mathbf{\rho }}=A \widetilde{\mathbf{\rho }}
\end{equation*}
where 
\begin{equation*}
A=\left( 
\begin{array}{ccc}
\frac{\partial H_{1}}{\partial \widetilde{\rho }_{1}}|_{\widetilde{\rho }=0}
& \ldots & \frac{\partial H_{1}}{\partial \widetilde{\rho }_{m}}|_{%
\widetilde{\rho }=0} \\ 
\vdots & \ldots & \vdots \\ 
\frac{\partial H_{m}}{\partial \widetilde{\rho }_{1}}|_{\widetilde{\rho }=0}
& \ldots & \frac{\partial H_{m}}{\partial \widetilde{\rho }_{m}}|_{%
\widetilde{\rho }=0}%
\end{array}%
\right) \text{.}
\end{equation*}

As we show in the appendix, filling in for the expressions of $\frac{%
\partial H_{i}}{\partial \widetilde{\rho }_{j}}|_{\widetilde{\rho }=0}$, we
can rewrite $A$ as 
\begin{equation*}
A=\left( 
\begin{array}{ccc}
\pi _{11}x_{1} & \ldots & \pi _{1m}x_{m} \\ 
\vdots & \ldots & \vdots \\ 
\pi _{m1}x_{1} & \ldots & \pi _{mm}x_{m}%
\end{array}%
\right)
\end{equation*}%
where 
\begin{equation*}
x_{i}=\dsum_d P_{i}(d)w_{i}(d)d\frac{f_{i}(d,1)}{g_{i}(d,0)}\text{.}
\end{equation*}

The term $x_i$ is a nicely interpretable factor. It is the relative growth
in infection for due to type $i$, but adjusted by the relative rates at
which type $i$'s will be met by other agents (so weighted by degrees
according to $w_i(d)$).

Note that if when we start with some vector of $\rho _{j}$'s near $0$ (so
our approximation is correct), but with positive entries, and then we end up
with a new vector that is at least as large as the starting vector, then it
must be that $0$ is an \textsl{unstable} solution.

\begin{definition}
There is diffusion from a small seed if and only if for any small $\varepsilon>0$,  
there exists some $\mathbf{v}$ such that $0<v_i<\varepsilon$ for all $i$ and
$A\mathbf{v}> \mathbf{v}$.
\end{definition}

Thus, diffusion from a small seed requires that beginning any small fraction
of initial adopters the ``dynamics'' lead to a larger fraction of adopters.

We remark that if 0 is unstable relative to some small initial seed $\mathbf{%
v}>0$, then it is unstable relative to any small initial seed $\widetilde{%
\mathbf{v}}>0$. That is, if $A \mathbf{v}>\mathbf{v}$, then for any $%
\widetilde{\mathbf{v}}>0$ there is some $t$ such that $A^t \widetilde{%
\mathbf{v}}> \widetilde{\mathbf{v}}$. Furthermore, if there is no diffusion
with a particular small initial distribution, then there will be no
diffusion with any other initial distribution. The next Lemma formalizes
such argument.\footnote{%
This result is partly an artifact of the continuous model approximation. For
an analysis of the importance of the specifics of initial adopters, see
Banerjee, Chandrasekhar, Duflo and Jackson (2011).}

\begin{lemma}
\label{lem-any} The condition for the diffusion from a small seed is
independent of the distribution across types of the initial seed. That is,
if $A \mathbf{v}>\mathbf{v}$ for some ${\mathbf{v}}>0$, then for any $%
\widetilde{\mathbf{v}}>0$ there is some $t$ such that $A^t \widetilde{%
\mathbf{v}}> \widetilde{\mathbf{v}}$.
\end{lemma}

\section{Analysis}

\subsection{Two Types}

We begin with the analysis of two types, which is a generalization of the
results in Section \ref{sectSIS}.

For now, we stick with a setting where $\pi_{11}=\pi_{22}=\pi$, so that
there is a symmetry in how introspective groups are in terms of their
meetings.

\begin{theorem}
\label{the-two} Let $\pi_{0}=\frac{1-x_{1}x_{2}}{{x_{1}+x_{2}}-2x_{1}x_{2}}$%
. Diffusion occurs if and only if one of the following conditions hold:

1) $x_{1}x_{2}>1$ or

2) $x_{1}x_{2}\leq 1$ and $\pi >\pi _{0}$.
\end{theorem}

The proof of Theorem \ref{the-two} appears in the Appendix. This result
generalizes what was found for the heterogeneous SIS model presented in
Section \ref{sectSIS}. The next corollary presents straightforward
consequences of it.

\begin{corollary}
\label{cor-two} In the two-type setting

1) If diffusion occurs within each type when isolated, then it would also
occur when there is interaction among the two.

2) If diffusion does not occur among either of the types when isolated, then
it would not occur when there is interaction among the two.

3) If diffusion would occur among only one of the types when isolated, then
it would occur among the entire population if homophily is high enough.
\end{corollary}

To see Corollary \ref{cor-two} first note that if there is only one type of
agent in the population then the condition for diffusion established by
Theorem \ref{the-two} reduces to the standard condition of $x>1$.
Therefore, diffusion occuring within each type when isolated corresponds to 
having $%
x_{1}>1$ and $x_{2}>1$. Those conditions in turn establish part 1) of the corollary as a
consequence of  part 1) of Theorem \ref{the-two}. If, on the contrary, diffusion does
not occur among either of the types when isolated, then $x_{1}<1$ and $x_{2}<1$.
Straightforward calculations show that then  the condition for diffusion stated in
part 2) of Theorem \ref{the-two} cannot satisfied for any value of $\pi \in
(0,1)$. The last part
of the corollary follows vacuously if $x_1x_2>1$, and otherwise diffusion occurs if $%
\pi$ exceeds $\pi_0$, establishing the claim.

\subsection{The General Case with Many Types}

Consider the following matrix $A$:

\begin{equation*}
A=\left( 
\begin{array}{ccc}
\pi _{11}x_{1} & \ldots & \pi _{1m}x_{m} \\ 
\vdots & \ldots & \vdots \\ 
\pi _{m1}x_{1} & \ldots & \pi _{mm}x_{m}%
\end{array}%
\right).
\end{equation*}

We remark that since $x_i>0$ for all $i$ (under our assumptions A1-A5), and
since $\Pi$ is primitive and nonnegative, it follows that $A$ is primitive
and thus $A^t>0$ for some $t$.

We can now state the following result, which generalizes the two-type result
to many types.

\begin{theorem}
\label{char} Diffusion occurs if and only if the largest eigenvalue of $A$
(denoted by $\mu $) is larger than $1$.
\end{theorem}

The proof of Theorem \ref{char} appears in the appendix.

Corollary \ref{cor-two} generalizes to the m-type case as presented next.

\begin{corollary}
\label{cor-char} 1) If diffusion from a small seed occurs within each type
when isolated, then it would also occur when there is interaction among
types.

2) If diffusion from a small seed does not occur for any of the types when
isolated, then it would not occur when there is interaction among them.

3) If there is some type for which $\pi_{ii}x_i>1$, then there is diffusion
from a small seed.

4) If there is a subset of types $S\subset T$ such that 
$\sum_{j\in S} \pi_{ij}x_j>1$ for each $i\in S$, then there is diffusion from a small seed.
\end{corollary}

We first explain why 1) holds, as 2) is a simple variation. If diffusion
occurs within each type when isolated then $x_{i}>1$ for all $i$ and
therefore 
\begin{equation*}
A>\left( 
\begin{array}{ccc}
\pi _{11} & \ldots & \pi _{1m} \\ 
\vdots & \ldots & \vdots \\ 
\pi _{m1} & \ldots & \pi _{mm}%
\end{array}%
\right).
\end{equation*}%
It follows that the largest eigenvalue of $A$ is larger than $1$ (since the
right-hand side matrix is a stochastic matrix and thus has a largest
eigenvalue of 1), and the result then follows from Theorem \ref{char}.

Next let us explain why 3) and 4) are true, and then discuss the intuition. 
3) is clearly a special case of 4), so let us discuss why 4) is true. 
Given that $\sum_{j\in S} \pi_{ij}x_j>1$
for each $i\in S$, it follows that for any positive vector $u$: $[Au]_i$ is greater than $\min_{j\in S} u_j$ for each $i\in S$.
Therefore, $\min_{j\in S} [Au]_j>\min_{j\in S} u_j$, and so it must 
be that if $u$ is the eigenvector corresponding to the maximum eigenvalue,\footnote{%
Again, recall that $A$ is primitive and thus has a strictly positive
eigenvector corresponding to its largest eigenvalue.} then
$Au>u$ and so the eigenvalue is larger than 1.  

1) and 2) of the corollary are fairly intuitive results. Note that
in case of just one population, then $x_{i}>1$ is the condition that
characterizes instability of (diffusion from) no activity. Thus, if all
populations are such that they would experience diffusion from a small seed
if isolated, then regardless of the interaction pattern there will be
diffusion; and similarly if none of them would experience diffusion in
isolation, then there cannot be diffusion when they interact.

The less obvious cases are 3) and 4), which show that
if some type or group of types has enough interaction with itself to get diffusion going,
then diffusion among the entire population will occur.  Again, these
emphasize the role of homophily in enabling diffusion (infection) from a
small seed:  if there is some
group of types that interacts within itself in a manner sufficient to enable
diffusion among that group, then a toehold can be established and diffusion
will occur from a small seed.

Another corollary is that if
populations are similar so that they have the same infection properties near
0 (i.e., $x_{i}=x_{j}=x$ for all $i$ and $j$), then diffusion properties are
determined by whether this growth rate is bigger or smaller than 1.  

\begin{corollary}
\label{cor-similar} If $x_{i}=x_{j}=x$ for all $i$ and $j$, then there is
diffusion from a small seed if and only if $x>1$.
\end{corollary}

This corollary then emphasizes that in order for the homophily and particular patterns
of interaction to matter, it must be that types are not just heterogeneous in their
interaction (the $\Pi$ matrix), but also in their adoption/infection proclivities.  If
they all have similar adoption/infection proclivities, then the particular details of 
who interacts with whom do not affect diffusion from a small seed.

The proof of this corollary is straightforward. Note that 
\begin{equation*}
A=x\Pi =x\left( 
\begin{array}{ccc}
\pi _{11} & \ldots & \pi _{1m} \\ 
\vdots & \ldots & \vdots \\ 
\pi _{m1} & \ldots & \pi _{mm}%
\end{array}%
\right) \text{.}
\end{equation*}%
It follows that the largest eigenvalue of $A$ is larger than $1$ if and only
if $x>1$ since $\Pi $ is a stochastic matrix and has a maximum eigenvalue of 
$1$.

The less obvious cases are thus such that there are some types who would
experience diffusion on their own, while others would not. Then the
interaction patterns really matter and, as already illustrated for the
two-type case, some subtle conditions ensue. A sufficient condition again is
that there is sufficient homophily such that infection can take hold within
some type, and then it can spread among the population, but more complicated
patterns among a number of groups can also possibly lead to diffusion from a
small seed.

\section{Concluding Remarks}

The focus of most of the related literature has been on analyzing the effect
that the degree distribution has on diffusion in social networks (see e.g.,
Jackson and Rogers, 2007, L\'{o}pez-Pintado, 2008, Galeotti and Goyal, 2009,
Galeotti et al., 2010.). This paper, however, focuses on the effect of
homophily, something which despite its importance has received little
attention in the diffusion literature. One of the few exceptions is the
paper by Golub and Jackson (2010) which also studies the impact of homophily
on some (very different) learning and diffusion processes. There are
important differences between our approach and theirs. On the one hand, the
diffusion processes analyzed are not the same; we focus on what can be
thought of as generalizations of the SIS infection model, whereas Golub and
Jackson (2010) analyze models of diffusion based either on shortest paths
communication, random walks or linear updating processes. Second, the paper
by Golub and Jackson (2010) studies the convergence time to the steady
state, whereas we analyze whether there is or not convergence to a state
with a positive fraction of adopters.

As a first step to understanding the effect of homophily on diffusion, in
this paper we have concentrated on a specific question; namely the spreading
of a new behavior when starting with a small initial seed. A central insight
here is that homophily can facilitate infection or contagion.

Nevertheless, there are other issues which are left for further work. For
example, one could evaluate the size of the adoption endemic state as a
function of the homophily level. There homophily might have conflicting
effects: although it can facilitate an initial infection, it might be that
an increase in homophily can also lead to a decrease in the overall
infection rate. Indeed, the eventual fraction of adopters attained in the
steady state might depend on the homophily level in complicated ways.

\section*{Appendix}

\noindent \textbf{Proof of Theorem \ref{the-one}}: The proof of Theorem \ref%
{the-one} is a straightforward consequence of the proof of Theorem \ref%
{the-two} as seen by substituting the functions $f_{i}(d,a)=\nu_i a$, $%
g_{i}(d,a)=\delta_i $ and $w_{i}(d)=\frac{d}{\langle d\rangle _{i}}$ and
obtaining the corresponding $x_i$'s.\hbox{\hskip3pt\vrule width4pt height8pt
depth1.5pt} \smallskip

\noindent \textbf{Proof of Theorem \ref{char}}: First, note that the system
of equations described describing the steady state is 
\begin{equation}
\widetilde{\rho }_{i}=H_{i}(\widetilde{\rho }_{1},\widetilde{\rho }_{2},...,%
\widetilde{\rho }_{m}),
\end{equation}%
where 
\begin{equation*}
H_{i}(\widetilde{\rho }_{1},\widetilde{\rho }_{2},...,\widetilde{\rho }%
_{m})=\sum\limits_{j}\pi _{ij}\sum\limits_{d}P_{j}(d)w_{j}(d)\frac{%
rate_{j,d}^{0\rightarrow 1}}{rate_{j,d}^{0\rightarrow
1}+rate_{j,d}^{1\rightarrow 0}}
\end{equation*}%
for $i\in \{1,...,m\}$.

This is approximated by a linear system in the neighborhood of $\mathbf{(}%
\widetilde{\rho }_{1},\ldots,\widetilde{\rho }_{n}\mathbf{)}=(0,\ldots,0)$
as follows: 
\begin{equation*}
\widetilde{\mathbf{\rho }}=A\widetilde{\mathbf{\rho }}
\end{equation*}%
where

\begin{equation*}
A=\left( 
\begin{array}{ccc}
\frac{\partial H_{1}}{\partial \widetilde{\rho }_{1}}|_{\widetilde{\rho }=0}
& \ldots & \frac{\partial H_{1}}{\partial \widetilde{\rho }_{m}}|_{%
\widetilde{\rho }=0} \\ 
\vdots & \ldots & \vdots \\ 
\frac{\partial H_{m}}{\partial \widetilde{\rho }_{1}}|_{\widetilde{\rho }=0}
& \ldots & \frac{\partial H_{m}}{\partial \widetilde{\rho }_{m}}|_{%
\widetilde{\rho }=0}%
\end{array}%
\right)
\end{equation*}

\noindent Note that%
\begin{equation*}
\frac{\partial rate_{i,d}^{0\rightarrow 1}}{\partial \widetilde{\rho }_{i}}%
=\sum_{a=0}^{d}f_{i}(d,a)\left( _{a}^{d}\right) \left( a\widetilde{\rho }%
_{i}{}^{a-1}(1-\widetilde{\rho }_{i})^{(d-a)}+(d-a)\widetilde{\rho }%
_{i}{}^{a}(1-\widetilde{\rho }_{i})^{(d-a-1)}\right)
\end{equation*}%
and therefore 
\begin{equation*}
\frac{\partial rate_{i,d}^{0\rightarrow 1}}{\partial \widetilde{\rho }_{i}}%
|_{0}=f_{i}(d,0)\left( _{0}^{d}\right) (d-0)(1-0)^{(d-1)}+f_{i}(d,1)\left(
_{1}^{d}\right) (1-0)^{(d-1)}=d\left[ f_{i}(d,1)+f_{i}(d,0)\right] \text{.}
\end{equation*}

\noindent Analogously 
\begin{equation*}
\frac{\partial rate_{i,d}^{1\rightarrow 0}}{\partial \widetilde{\rho }_{i}}%
|_{0}=d\left[ g_{i}(d,1)+g_{i}(d,0)\right] \text{.}
\end{equation*}

\noindent Then%
\begin{equation*}
\frac{\partial H_{i}}{\partial \widetilde{\rho }_{j}}|_{\widetilde{\rho }%
=0}=\pi _{ij}\sum\limits_{d}P_{j}(d)w_{j}(d)\frac{\frac{\partial
rate_{j,d}^{0\rightarrow 1}}{\partial \widetilde{\rho }_{j}}%
|_{0}rate_{j,d}^{1\rightarrow 0}|_{0}-rate_{j,d}^{0\rightarrow 1}|_{0}\frac{%
\partial rate_{j,d}^{1\rightarrow 0}}{\partial \widetilde{\rho }_{j}}|_{0}}{%
\left( rate_{j,d}^{0\rightarrow 1}+rate_{j,d}^{1\rightarrow 0}\right)
^{2}|_{0}}
\end{equation*}%
and thus, 
\begin{equation*}
A=\left( 
\begin{array}{ccc}
\pi _{11}x_{1} & \ldots & \pi _{1m}x_{m} \\ 
\vdots & \ldots & \vdots \\ 
\pi _{m1}x_{1} & \ldots & \pi _{mm}x_{m}%
\end{array}%
\right)
\end{equation*}%
where 
\begin{equation*}
x_{i}=\dsum\limits_{d}P_{i}(d)w_{i}(d)d\frac{%
f_{i}(d,1)g_{i}(d,0)-f_{i}(d,0)g_{i}(d,1)}{(f_{i}(d,0)+g_{i}(d,0))^{2}}\text{%
. }
\end{equation*}

\noindent Given A1, $x_{i}$ can be rewritten as 
\begin{equation*}
x_{i}=\dsum P_{i}(d)w_{i}(d)d\frac{f_{i}(d,1)}{g_{i}(d,0)},
\end{equation*}%
which is well defined since A4 holds. 



\noindent As mentioned in the text, $A$ is primitive since $\Pi$ is
primitive and since A1 and A4 are satisfied implying that $x_i>0$.\footnote{%
In fact, with two types $A$ is a positive matrix since $0<\pi <1$.} Thus, by
the Perron--Frobenius Theorem (which applies to primitive matrices) the
maximum eigenvalue, denoted $\mu $ hereafter, is positive and its
corresponding eigenvector, denoted by $\mathbf{u}$ hereafter, is also
positive.

We show next that the condition for diffusion from a small seed, or the
instability of $\widetilde{\rho }=0$, corresponds with the condition that
the largest eigenvalue of $A$ is larger than $1$. 

Let us first show that if $\mu >1$ then $\widetilde{\rho }=0$ is
unstable. Note that if $\mu >1$ then 
\begin{equation*}
A\delta \mathbf{u}=\mu \delta \mathbf{u}>\delta \mathbf{u}.
\end{equation*}%
Thus, picking small enough $\delta $ so that $\delta u_{i}<\varepsilon $ for
each $i$, satisfies the definition of diffusion from a small seed with $%
\delta \mathbf{u}$ (or instability of 0).

To see the converse, first consider the case such that $\mu < 1$.
Given $\varepsilon >0$ consider any $\mathbf{v}$ such that $%
0<v_{i}<\varepsilon $ for all $i$. Suppose that $A\mathbf{v}>\mathbf{v}$. It
then follows $A(A\mathbf{v})>A\mathbf{v}>\mathbf{v}$ as $A$ is nonnegative
and has at least one positive entry in each row. Iterating, it follows that
that $A^{t}\mathbf{v}>\mathbf{v}$ for any $t$. However, choose $\delta $
such that $\delta \mathbf{u}>\mathbf{v}$. Given that $A$ is is nonnegative
and has at least one positive entry in each row, and both vectors are
positive, it follows that $A\delta \mathbf{u}>A\mathbf{v}$, and similarly
that 
\begin{equation*}
A^{t}\delta \mathbf{u}>A^{t}\mathbf{v}.
\end{equation*}
Given our previous claim, this then implies that 
\begin{equation*}
A^{t}\delta \mathbf{u}>\mathbf{v}
\end{equation*}%
for all $t$. However, 
\begin{equation*}
A^{t}\delta \mathbf{u}=\delta \mu ^{t}\mathbf{u}\rightarrow 0
\end{equation*}%
given that $\mu <1$, which is a contradiction.

To complete this part of the proof consider the case such that $%
\mu =1 $. Consider $\varepsilon >0$. Consider any vector $\mathbf{v}$ such
that $\mathbf{v}_{i}<\varepsilon $. Note that for any small enough $\delta>0$
the largest eigenvalue of $A-\delta I$ is less than 1. Thus, by the argument
above, $(A-\delta I)\mathbf{v}$ is not greater than $\mathbf{v}$. Therefore, 
$A\mathbf{v}$ is not greater than $\mathbf{v}$.\hbox{\hskip3pt\vrule
width4pt height8pt depth1.5pt}

\medskip

\noindent \textbf{Proof of Theorem \ref{the-two}}: We have already shown
that $\widetilde{\rho }=0$ is unstable if and only if the largest eigenvalue
of matrix $A$ is above $1$. Let us now complete the proof by examining the
eigenvalue in the two-type case. The eigenvalues of a $2\times 2$ matrix are
easily computed. Writing 
\begin{equation*}
A=\left( 
\begin{array}{cc}
a_{11} & a_{12} \\ 
a_{21} & a_{22}%
\end{array}%
\right) \text{,}
\end{equation*}%
the largest eigenvalue of $A$ is\footnote{%
Note that since $A$ is primitive, its largest eigenvalue is real and
positive.} 
\begin{equation*}
\mu =\frac{(a_{11}+a_{22})+\sqrt{%
(a_{11}+a_{22})^{2}-4(a_{11}a_{22}-a_{12}a_{21})}}{2}
\end{equation*}%
or equivalently 
\begin{equation*}
\mu =\frac{a_{11}+a_{22}+\sqrt{%
(2-a_{11}-a_{22})^{2}-4+4a_{11}+4a_{22}-4a_{11}a_{22}+4a_{12}a_{21}}}{2}%
\text{.}
\end{equation*}%
Thus, $\mu $ is larger than $1$ if and only if 
\begin{equation}
\frac{a_{11}+a_{22}}{2}>1  \label{cond 1}
\end{equation}%
or 
\begin{equation}
-1+a_{11}+a_{22}-a_{11}a_{22}+a_{12}a_{21}>0.  \label{cond 2}
\end{equation}

\noindent Given that $a_{11}=\pi x_{1}$, $a_{22}=\pi x_{2}$, $a_{12}=(1-\pi
)x_{2}$ and $a_{21}=(1-\pi )x_{1}$ then conditions (\ref{cond 1}) and (\ref%
{cond 2}) imply that diffusion (i.e., instability of $\widetilde{\rho }=0$)
occurs if and only if 
\begin{equation}
\pi >\frac{2}{x_{1}+x_{2}}
\end{equation}%
or 
\begin{equation}  \label{twoprime}
\pi (x_{1}+x_{2}-2x_{1}x_{2})+x_{1}x_{2}-1>0\text{.}
\end{equation}

\noindent \textbf{Case 1:} $\frac{x_{1}+x_{2}}{2x_{1}x_{2}}>1$. In this
case, condition (\ref{twoprime}) is equivalent to 
\begin{equation*}
\pi >\frac{1-x_{1}x_{2}}{x_{1}+x_{2}-2x_{1}x_{2}}
\end{equation*}%
and therefore diffusion occurs in this case if and only if%
\begin{equation}  \label{greater}
\pi >\min \{\frac{1-x_{1}x_{2}}{x_{1}+x_{2}-2x_{1}x_{2}},\frac{2}{x_{1}+x_{2}%
}\}.
\end{equation}

\noindent \textbf{Case 2:} $\frac{x_{1}+x_{2}}{2x_{1}x_{2}}<1$. In this
case, condition (\ref{twoprime}) is equivalent to 
\begin{equation*}
\pi <\frac{x_{1}x_{2}-1}{2x_{1}x_{2}-x_{1}-x_{2}}
\end{equation*}%
and therefore diffusion occurs in this case if and only if%
\begin{equation}  \label{less}
\frac{2}{x_{1}+x_{2}}<\pi \ \mathrm{or } \ \pi<\frac{x_{1}x_{2}-1}{%
2x_{1}x_{2}-x_{1}-x_{2}}\text{.}
\end{equation}

\noindent \textbf{Case 3:} $\frac{x_{1}+x_{2}}{2x_{1}x_{2}}=1$. In this
case, condition (\ref{twoprime}) simplifies to $x_1x_2>1$, and and therefore
diffusion occurs in this case if and only if%
\begin{equation}  \label{equal}
\frac{2}{x_{1}+x_{2}}<\pi \ \mathrm{or} \ x_1x_2>1
\end{equation}

Let us now show part (1) of Theorem \ref{the-two}.

\begin{figure}[ht]
\begin{center}
\includegraphics[height=3.5in]{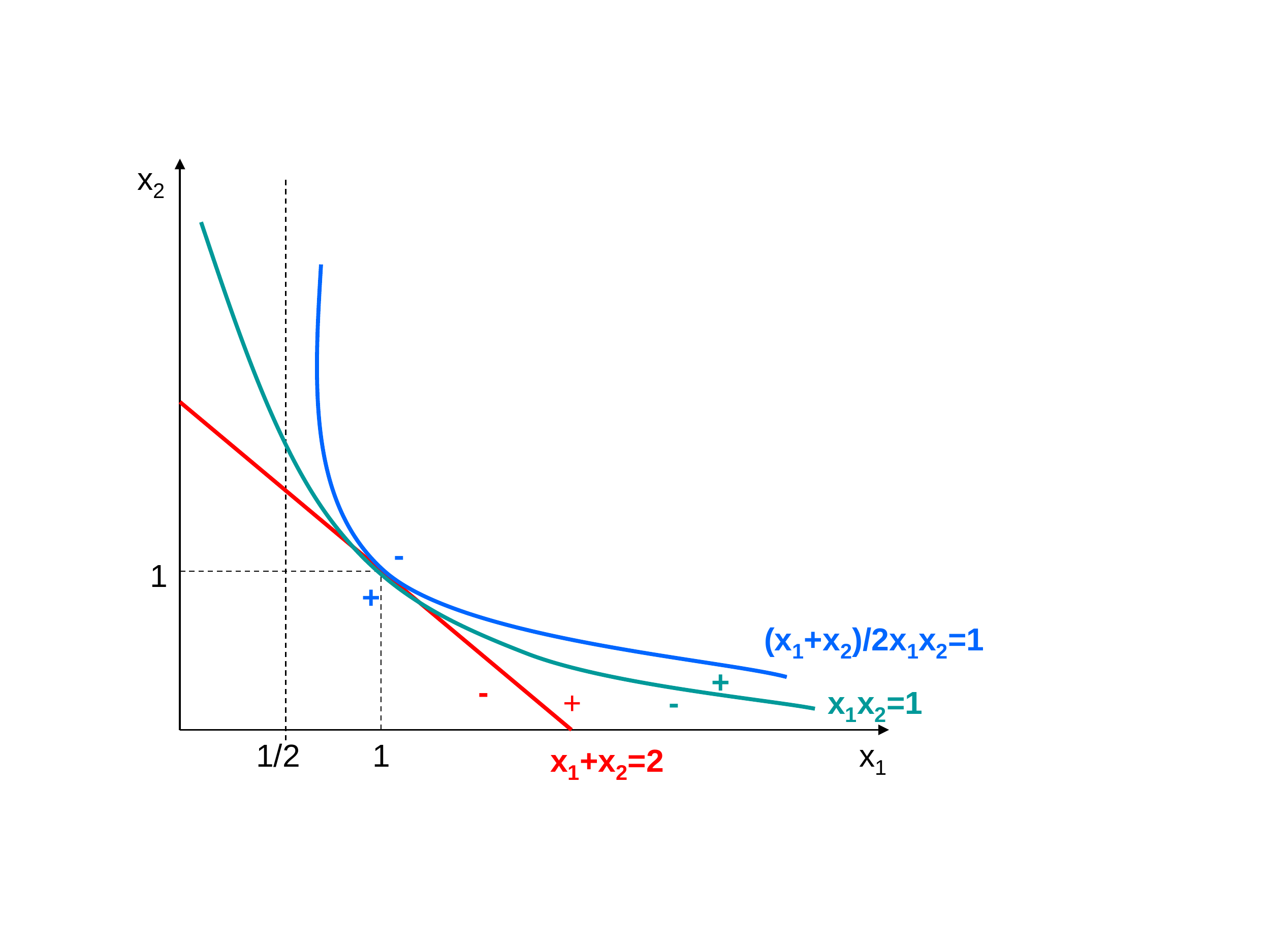}
\end{center}
\caption{The relationship between the key expressions in the proof of
Theorem \protect\ref{the-two}.}
\label{fig1}
\end{figure}

Suppose that $x_{1}x_{2}>1$ holds. Then $\frac{x_{1}+x_{2}}{2x_{1}x_{2}}$
can fall into any of the cases above. If it were greater than 1, then $\frac{%
1-x_{1}x_{2}}{x_{1}+x_{2}-2x_{1}x_{2}}<0$ which in particular by Case 1 and (%
\ref{greater}) implies that there is diffusion for any $\pi \in (0,1)$. If
it were equal to 1, then by Case 3, the result holds. If instead $\frac{%
x_{1}+x_{2}}{2x_{1}x_{2}}<1$ then Case 2 applies. In that case, referring to
Figure \ref{fig1}, $(x_1, x_2)$ lies above the upper-most curve,\footnote{%
The relative positions of the curves are easily checked, and note the plus
and minus signs that indicate whether one is above or below 1 for the
corresponding colored expression.} and it is clear that there would exist
another profile $(\widehat{x}_{1},\widehat{x}_{2})$ such that $\widehat{x}%
_{1}\leq x_{1}$ and $\widehat{x}_{2}\leq x_{2}$ and which lies in the
regions considered previously (that is, where $\frac{1-x_{1}x_{2}}{%
x_{1}+x_{2}-2x_{1}x_{2}}\leq 0$). Therefore diffusion for $(\widehat{x}_{1},%
\widehat{x}_{2})$ occurs for all $\pi \in (0,1)$, which in particular
implies that for the larger case $(x_{1},x_{2})$ diffusion would also occur
for all $\pi \in (0,1)$ as the largest eigenvalue of a larger matrix is
necessarily larger than the largest eigenvalue of a smaller matrix.

Next, we show part (2) of Theorem \ref{the-two}. Suppose that $%
x_{1}x_{2}\leq 1$. This implies that $\frac{x_{1}+x_{2}}{2x_{1}x_{2}}>1$
(see Figure \ref{fig1}) or else that $x_1=x_2=1$ in which Case 3 applies and
there cannot be diffusion. Thus, let us analyze the situation where $\frac{%
x_{1}+x_{2}}{2x_{1}x_{2}}>1$ and Case 1 applies. Diffusion occurs if and
only if $\pi >\min \{\frac{1-x_{1}x_{2}}{x_{1}+x_{2}-2x_{1}x_{2}},\frac{2}{%
x_{1}+x_{2}}\}$. Note that if $x_{1}+x_{2}<2$ then $\frac{2}{x_{1}+x_{2}}>1 $
and therefore diffusion occurs if and only if $\pi >\frac{1-x_{1}x_{2}}{%
x_{1}+x_{2}-2x_{1}x_{2}}$. If, on the contrary, $x_{1}+x_{2}\geq 2$ then, it
is straightforward to show that $\frac{2}{x_{1}+x_{2}}>\frac{1-x_{1}x_{2}}{%
x_{1}+x_{2}-2x_{1}x_{2}}$ which also implies that diffusion in such a case
occurs if and only if $\pi >\frac{1-x_{1}x_{2}}{x_{1}+x_{2}-2x_{1}x_{2}}$.%
\hbox{\hskip3pt\vrule width4pt height8pt depth1.5pt}

\medskip

\noindent \textbf{Proof of Lemma \ref{lem-any}:} Given the proof of Theorem %
\ref{char}, it follows that if $A\mathbf{v}>\mathbf{v}$ for some ${\mathbf{v}%
}>0$ then $\mu >1$. Then, choose $\delta $ such that $\delta \mathbf{u}<%
\widehat{\mathbf{v}}$. It follows that $A\delta \mathbf{u}<A\widehat{\mathbf{%
v}}$ (since $A$ is nonnegative and has at least one positive entry in each
row), and similarly that 
\begin{equation*}
\mu ^{t}\delta \mathbf{u}=A^{t}\delta \mathbf{u}<A^{t}\widehat{\mathbf{v}},
\end{equation*}%
and the first expression is growing with $\mu ^{t}$.%
\hbox{\hskip3pt\vrule
width4pt height8pt depth1.5pt}

\end{document}